# Interaction of Economic Freedom and Foreign Direct Investment Globally: Special Cases from Neglected Regions

**Yhlas SOVBETOV**[a*], **Mohamed MOUSSA**[b]

[a] *Department of Economics, London School of Commerce, United Kingdom*

[b] *Department of Economics, Istanbul University, Turkey*

## Abstract

*This paper studies the macroeconomic impact of economic freedom on foreign direct investments inflows in both global and regional panel analyses involving 156 countries through the period of 1995-2016. Unlike to prior literature, it includes often neglected nations such as Fragile and Conflict-Affected states, Sub-Saharan, Oceanian, and Post-Soviet countries. The paper finds a positive impact of economic freedom on FDI under fixed-effects model in global case where a unit change in economic freedom scales FDI inflows up to 1.15 units. More specifically, all 9 regions also refer to positive and significant impact of economic freedom on FDI. The highest impact is recorded in European countries, whereas the lowest ones are documented in Fragile-Conflict affected states, Sub-Saharan zone, and Oceanian countries.*

* Corresponding author. Office CH-28, London School of Commerce, Chaucer House, London, SE1 1NX, London, UK. Tel: +44 207 357 0077 / ext.364
E-mail addresses: ihlas.sovbetov@lsclondon.co.uk (Y.Sovbetov), medmoussane@yahoo.fr (M. Moussa)

## 1. Introduction

The latest version of economic growth model (Eduardo Borensztein, Jose De Gregorio, and Jong-Wha Lee, 1998) advocates that host countries can inherit from foreign direct investment (FDI) in many ways. First, it contributes to growth through capital accumulation by attracting money flows into the country. The second way is through training and development of local citizens in order to get them used to the new technology. Besides, FDI also increases the level of competition and improves resource allocation by enhancing domestic financial market and lowering the cost of acquiring new capital. It is therefore imperative and necessary for countries to pin down policies that favor and attract more investment in order to better the standard of living of their population.

In fact, Kusi Hornberger, Joseph Battat, and Peter Kusek (2011) refers that FDI has increased globally not only in developed countries but also in least developed and transition economies. More specifically they state that their share in global FDI has scaled from 19% in 2000 up to 52% in 2010, hence it estimates that the volume of global FDI will attain to level of $3-4 trillion in 2014. Indeed FDI has increased due to recent amplifications in links between nations through internationalization process, as well as due to improvements in economic freedom of developed and emerging markets. Even a number of recent studies (Elizabeth Asiedu 2006; Kazeem Bello Ajide and Perekunah Bright Eregha 2014; Bosede Victoria Kudaisi 2014) find that most of Sub-Saharan countries have become more attractive in terms of FDI inflows, but its share in global FDI is less than 2% (Mory Fode Fofana 2014). On contrary, FDI of Arab countries has negatively affected by recent instabilities and conflicts, i.e. Arab Spring, arose in the Middle East. Indeed, to our knowledge, these countries are often marginalized and excluded from studies which emerge a gap in literature.

In this study we attempt to fill this gap by including these fragile-conflict affected states as well as often neglected Post-Soviet countries in our sample to examine the impact of economic freedom on FDI. Beside the global analysis of 156 countries through the period of 1995-2013, the study also gives a sight to the region-based interaction of FDI inflows with economic freedom level of nations. Our results reveal that a 1% increase in economic freedom triggers the global FDI up by 6%. Furthermore, the aftermath of region-based analysis indicates that a percentage change in economic freedom alters FDI inflows by 17% in Europe, 12% in Asia, 7% in Africa, and 8-9% in North and South America. Additionally the analysis of neglected regions also shows significant results where 1% increase in economic freedom boosts FDI inflows by 14% in Sub-Saharan Africa, 10% in Post-Soviet countries, 5% in Fragile-Conflict affected states, and 4% in Oceania region.

## 2. Literature Review

A sizable empirical literature exists on macroeconomic impact of economic freedom and its components on FDI. For instance, Rakesh Sambharya and Abdul Rasheed (2015) examine the macroeconomic effects of economic and political freedom on FDI inflows in 95 host countries in a panel data analysis through the periods of 1995-2000. Their results suggest before benefiting from FDI inflows, countries need to emphasize on a better economic management in terms sound monetary policy, fiscal burden, and banking and finance. Additionally they advocate that less government participation into an economy, strong property rights, low prevalence of informal markets, and less corruption are desirable for more FDI inflows.

On the other hand, Dennis Pearson, Dong Nyonna, and Kil-Joong Kim (2012) investigate the impact of economic freedom and growth on FDI in state levels, indifferent to most studies that consider determinants of FDI inflows into United States as a country. They use a panel data analysis of 50 states through the period of 1984-2007 employing random-effects model. They find that both growth and economic freedom have significant positive impact on FDI in all states. However, the authors also explore that per capita income and unemployment rate cause negative impact on FDI. They address these relations to the fact that states with higher per capita income repel FDI inflows since higher income implies higher wages, and high unemployment rate is positively associated with crime ratio, thus discourages investors' interests.

Likewise Mart Bengoa and Blanca Sanchez-Robles (2003) also examine the interplay between economic freedom, growth, and FDI inflows using a panel data analysis of sample of 18 Latin American countries from 1970-1999. They observe that economic freedom remains positive and significant both in fixed- (0.0043) and random-effects (0.0046) regression models deriving similar coefficient magnitudes which imply their robustness. On the other hand, the impact of growth on FDI appears significant only in fixed-effects model with magnitude of 0.01. Eventually, they conclude that both economic freedom and growth in host countries generate benefits on FDI inflows only if there is a sufficient human capital accompanied by economic stability and liberalized markets.

Furthermore, Asiedu (2006) studies the role of natural resources (export of oil, gold and others), government policy (human capital in terms of literacy rate, quality of infrastructure, and inflation rate), market size (income per capita), institutions (rate of corruption and rule of law) and political instability (number of coup, assignations and revolutions) on FDI in a panel data analysis of 22 African countries from 1984 to 2000. She employs Hausman test and finds that the random-effects model generates biased estimators. Preferring the fixed-effects model she exhibits that a unit change in openness of economy alters FDI by 0.20 units when policy variable is proxied with human capital (literacy rate), and by 0.23 units when it is proxied with infrastructure investments (landline phone penetration) of the country. Here, she specifies that an increase in FDI does not always indicate amplification in economic growth, because she addresses an ambiguous empirical relation of these two in literature as some studies that stipulate augmentations of economic growth with certain conditions such as when the host country has higher quality education (Borensztein et al., 1998); optimal income level (Matthew Tyler Lund, 2010); or well-established financial markets and regulations (Maria Carkovic and Ross Levine, 2005).

Besides, Fofana (2014) measures the influence of economic freedom components on FDI in 25 Western European and 26 Sub-Saharan countries through 2001-2009 where he discovers that the aggregate index of economic freedom is not a significant explanatory of FDI for African case, but European countries. He proxies economic freedom with three institutional variables such as "the size of the economy", "the size of the population", and "the legal system and rule of law"; and with three regulatory variables such as "size of government", "freedom of international trade", and "regulations of labor, credit, and business". As a results he observes that only "legal system and rule of law" variable appears significant in African sample, where it fails to be significant in European sample. More specifically, the author also discovers positive links between GDP and FDI, and Population and FDI; meanwhile he finds negative association of Natural Resources and FDI in fixed-effects model with cross-section dummy variables where he accounts 94% of variation in FDI. He addresses it to the current stage of this region which is in the development process. Besides he also admits that insufficient observation number is another restriction of his study which leads to insignificant results.

Nonetheless, he finds plausible results for European sample where economic freedom, i.e. that is proxied by *"size of government"*, *"freedom of international trade"*, and *"regulations of labor, credit, and business"*, appears statistically significant determinant of FDI. He also explores very similar results as African case with fixed-effects model that includes cross-section dummy variables. The only difference between cases appears as disappearance of significant impact of population on FDI in European sample.

On the other hand, Rahim Quazi (2007) investigates the collision of economic independence on the flow of foreign investment in a panel data regression for seven major East Asian countries over 1995-2000 periods, employing both fixed-and random-effects models. Initially he examines the full sample where 70% of FDI is explained by its first lag, political instability, and market size variables in random-effects model. But both in random-effects and GLS models the economic freedom fails to be significant.

However when he adds a dummy variable for China, a country in sample that requires a exceptional attention due to being magnet for FDI, both random-effects and GLS models estimate significant but negative impact of economic freedom on FDI. Indeed this negative coefficient implies positive impact on FDI. Because he proxies the economic freedom variable with domestic investment climate that is constructed on a scale of 1 to 5, where 1 indicates set of policies most favorable to economic freedom, and 5 represents policies with least conducive. The outcomes unveil that the dummy variable tends to be very significant with a magnitude of 3.07 in the fixed panel and 3.43 in the random panel. In addition all other explanatory variables (change in the volume of FDI inflows, political stability, market size and level of profit in investment) turn out to be significant except quality of infrastructure and human capital. He concludes that investment flows to China more than other countries in the sample because of the huge natural resources, its cheap labor cost, also the geographic proximity to Hong Kong and Taiwan, the recent forms in the economic sector are also other factors.

Moreover when he considers taking the China out of the sample, he finds quite similar (a bit larger) negative and significant coefficient for economic freedom as the case of dummy variable, and makes similar interpretations. But he emphatically states that the sample countries still encounter regional bias in terms of FDI which is definitely favorable for China.

In another study Sufian Eltayeb Mohamed and Moise Sidiropoulos (2010) look at the determinants of FDI in 12 MENA (Middle East North African countries) where their find in line results with the traditional literature of economic freedom and FDI. To capture more variations in FDI, they include domestic, financial, institution, policy, and other external variables into fixed-effects model, and compare estimations of MENA countries with other developed ones. They proxy domestic factors with market size (logarithm of GDP); financial factors with national stock index; institutional factors with investment profile and corruption levels; policy factors with inflation rate and government spending; and external factors with global liquidity and trade freedom. As a result they find out that the FDI is largely determined by the market size and trade freedom which generate coefficient of 98.15 and 12.43, alongside with minor determinants such as investment profile, corruption level, inflation rates, government spending, natural resources, and growth expectation. Unlike to these results, in case of MENA countries the trade freedom turns out insignificant. Indeed it might be due to political instabilities and conflicts in this region. Latterly, Bounoua Chaib and Matallah Siham (2014) also address to the same issues by referring importance of institutional quality and political stability in order to attract FDI in Algeria.

Lastly, all discussed literature studies are briefly summarized in the table 1.

**Table 1.** Summary of Empirical Studies

| Reference | Origin | Sample | Period | Model | Result |
|---|---|---|---|---|---|
| Sambharya & Rasheed (2015) | Global | 95 Countries | 1995-2000 | Panel Data | Government participation and corrupt level has negative impact on FDI, where secure property rights has positive. |
| Pearson et al. (2012) | US | 50 States | 1984-2007 | Panel Data | -Growth and economic freedom has positive impact on FDI.<br>-Per capita income and unemployment rate have negative impact on FDI. |
| Bengoa & Sanchez-Robles (2003) | Latin America | 18 Countries | 1970-1999 | Panel Data | Economic freedom and growth generate FDI inflows only if a country has sufficient human capital accompanied by economic stability and liberalized markets. |
| Asiedu (2006) | Africa | 22 Countries | 1984-2000 | Panel Data | -Economic growth does not always increases FDI.<br>-A unit change in economic freedom alters FDI by 0.20 units when policy variable is proxied with human capital, and by 0.23 units when it is proxied with infrastructure investments. |
| Fofana (2014) | EU, Sub-Saharan | 25 West EU, 26 Sub-Saharan Countries | 2001-2009 | Panel Data | -Legal system and rule of law is significant determinant of FDI in Sub-Saharan states.<br>-GDP has positive, and Natural Resources have negative impact on FDI in both West EU and Sub-Saharan states. |
| Quazi (2007) | East Asia | 7 Countries | 1995-2000 | Panel Data | -Including China into the sample makes impact of economic freedom on FDI insignificant.<br>-After controlling China with dummy variable, the impact turns to be significantly positive. |
| Mohamed & Sidiropoulos (2010) | MENA | 12 Countries | 1975-2006 | Panel Data | -The major determinants of global FDI are market size and trade freedom, whereas the minor determinants are investment profile, corruption level, inflation rates, government spending, and natural resources.<br>-For MENA countries trade freedom is insignificant. |

The rest of the paper is organized as follows. The next section describes the data and methodology of this study, and the results are reported in section 4. The final section includes concluding remarks about our analysis and its findings.

## 3. Data and Methodology

This study examines the macroeconomic impact of economic freedom on the foreign direct investment (FDI) inflows over the globe. The initial sample size was comprised of 189 countries over the period of 1995-2016. However due to unavailability of macro data for 33 countries, the sample size decreased to 156 countries.

The freedom of economic activity of the country is proxied by Economic Freedom Index (EDI) which is formed by Business Freedom Index (BFI), Trade Freedom Index (TFI), Investment Freedom Index (IFI), and Financial Freedom Index (FFI). The data for these indexes are gathered from online database of Heritage Foundation. We also investigate magnitude of FDI and EFI interaction on the regional basis holding the control variables such as GDP growth, Import and Export per GDP, Inflation, and Interest rates. The data for these variables are derived from online database of World Bank. Unlike to prior literature our study pursues the analysis with larger sample where often neglected nations such as fragile and conflict-affected states, sub-Saharan areas, and Oceania countries are also captured. Meantime with panel data analysis, we explore both fixed- and random-effects approaches, as well as a pooled regression of EFI on FDI.

**Table 2.** Descriptive Statistics of Data

| | *Mean* | *Median* | *Max.* | *Min.* | *Std. Dev.* | *Skewness* | *Kurtosis* | *N* |
|---|---|---|---|---|---|---|---|---|
| **FDI** | 0.8667 | 1.0486 | 6.0653 | -4.6052 | 1.3310 | -0.8586 | 4.8197 | 3432 |
| **Growth** | 1.1263 | 1.4110 | 5.0104 | -4.6052 | 1.1258 | -1.3923 | 5.5584 | 3432 |
| **Import** | 3.7035 | 3.6951 | 6.0517 | 2.1247 | 0.5064 | 0.2413 | 3.8322 | 3432 |
| **Export** | 3.5624 | 3.5907 | 5.4393 | 0.0000 | 0.5948 | -0.1278 | 3.7964 | 3432 |
| **Inflation** | 1.5231 | 1.6448 | 7.8748 | -4.6052 | 1.3996 | -0.3511 | 4.5987 | 3432 |
| **Interest** | 1.6828 | 1.6956 | 5.3151 | -4.6052 | 1.0732 | -1.3060 | 9.6637 | 3432 |
| **EFI** | 4.0463 | 4.0792 | 4.5054 | 2.3026 | 0.2407 | -1.8282 | 9.6284 | 3432 |

**Notes:** The log linearization technique is applied to data.

Following Bengoa and Sanchez-Robles (2003) approach, we extend their model by including macro control variables into the model as below.

$$FDI_{at} = \beta_0 + \beta_{1at}EFI_{at} + \sum_{j=2}^{6} \beta_{jat}M_{jat} + e_{at} \qquad \text{(Eq. 1)}$$

where *FDI* is foreign direct investment inflows of *country "a"* as percentage of its GDP at time *"t"*; *c* is a intercept; and *M* stands for five macro control variables of country *"a"* at *time "t"* respectively. Hence, *"e"* represents the residual term of the model.

To find out the best model for our panel data, we shall look to consistency and efficiency of GLS estimators through cross-section fixed (FE) and random effects (RE). Both of these models have potential advantages –as well as disadvantages– in their selection. The FE model assumes heterogeneity among all entities by allowing to have their own intercept values. However while this intercept differs among entities, it does not change over the time. Therefore FE model generates unbiased estimates of $\beta_i$, but it may suffer from high variance due to a larger variation between sample (country) to sample. In this case, our model with FE specification becomes as below.

$$FDI_{at} = \beta_0 + \sum_{i=1}^{156} \alpha_{ia}D_{ia} + \beta_{1at}EFI_{at} + \sum_{j=2}^{6} \beta_{jat}M_{jat} + e_{at} \qquad \text{(Eq. 2)}$$

where $D_a$ is a dummy variable which equates 1 for the country *"a"*, and zero for others in the sample. We also could include a fixed effect for period by considering a dummy variable for years as *"$D_t$"* only in case when the period is different for countries in the sample.

On the other hand, the RE model heals the high variance problem by generating estimates closer, on average, to the true value of any particular country as below.

$$FDI_{at} = \beta_{0a} + \beta_{1at}EFI_{at} + \sum_{j=2}^{6} \beta_{jat}M_{jat} + \varepsilon_{at} \qquad \text{(Eq. 3)}$$

$\beta_{0a} = \beta_0 + \omega_a$ where $\omega_a \sim N(0, \sigma^2)$

When the $\beta_{0a}$ is plugged into first equation model, it becomes as below.

$$FDI_{at} = \beta_0 + \beta_{1at}EFI_{at} + \sum_{j=2}^{6} \beta_{jat}M_{jat} + \omega_a + e_{at} \qquad \text{(Eq. 4)}$$

$$FDI_{at} = \beta_0 + \beta_{1at} EFI_{at} + \sum_{j=2}^{6} \beta_{jat} M_{jat} + u_{at} \qquad \text{(Eq. 5)}$$

where $u_{at} = \omega_a + e_{at}$. However, due to potential correlation between covariates of explanatory variables and $\omega_a$ the *($\beta_{1\text{-}7}$)* estimates of RE model are often biased. Unlike FE model, it captures both *"within"* and *"between"* deviations, and allows all entities to have a common mean value for intercept. With other words, the dummy variable *"$D_a$"* -was a part of intercept in the FE- becomes a part of error *"$e_a$"* in the RE model.

A prior to researcher's preference in trade-off between bias and variance, it is more logic to exhibit the dataset and characteristics of the sample. Besides there are few statistical tests that might be a guideline (table 3) in selection an appropriate model. According to this, initially two tests are employed: the Redundant Fixed Effects (RFE) and Breusch-Pagan Lagrange Multiplier (BP LM) tests to find out whether our panel data contain respectively a fixed effect and a random effect. In special case when both fixed and random effects are observed the Hausman test is recommended which is modeled as below.

$$H = (\beta_1 - \beta_0)'(Var(\beta_1) - Var(\beta_0))^{\rho}(\beta_1 - \beta_0) \qquad \text{(Eq. 6)}$$

where $\rho$ is pseudoinverse. The $H_0$ specifies that both $\beta_0$ (FE estimator) and $\beta_1$ (RE estimator) are consistent, but $\beta_1$ is efficient while $\beta_0$ is not. The alternative hypothesis indicates that only $\beta_0$ is consistent, and $\beta_1$ is not. However Andrew Bell and Kelvyn Jones (2015) criticize this analysis by stating that it is not a test of FE versus RE, but it is a test of the similarity of within and between effects. They assert that a RE model which accurately specifies the within and between effects will produce identical results to FE, regardless of the result of a Hausman test. They question the validity of FE model by accusing it as *"between effects, other higher-level variables and higher level residuals, none of which can be estimated with FE, should not be dismissed lightly; they are often enlightening, especially for meaningful entities such as countries. For these reasons, ..., RE models are the obvious choice"*.

**Table 3.** Fixed- and Random-Effects Model Selection

| Redundant Fixed Effect Test | Breusch-Pagan & Honda LM Tests | Concluded Model |
|---|---|---|
| $H_0$ is not rejected (No fixed effect) | $H_0$ is not rejected (No random effect) | Data are poolable (Pooled OLS) |
| $H_0$ is rejected (Fixed effect) | $H_0$ is not rejected (No random effect) | Fixed Effect Model (LSDV or GLS) |
| $H_0$ is not rejected (No fixed effect) | $H_0$ is rejected (Random effect) | Random Effect Model (GLS) |
| $H_0$ is rejected (Fixed effect) | $H_0$ is rejected (Random effect) | (1) Both Fixed & Random Effect Models<br>(2) Hausman Test (recommended) |

**Notes:** The null hypothesis for both Breush-Pagan and Honda LM tests is "No Random Effects". The null hypothesis for Redundant Fixed Effect test is "No Unobserved Heterogeneity (No Fixed Effect)".

In addition to those theoretical considerations, many researchers (Satkartar Kinney and David Dunson, 2007; Hun Myoung Park, 2009; Howard Bondell, Arun Krishna, and Sujit Gosh, 2011; Tom Clark and Drew Linzer, 2014; Bell and Jones, 2015) suggest to account practical and technical grounds in decision stage. They recommend evaluating the sample characteristics and objectives in trade-off between fixed and random-effects model selection. They argue that fixed-effects model makes sense under these 2 conditions. The first, if all entities (groups) in the sample are functionally identical. Second, if the goal is to assess common effect magnitude only for sampled entities, but not to generalize it to other entities.

On this basis, we assume that the model which would represent our panel data is more likely to be RE, as countries (entities) in the sample are not functionally identical, and this empirical study aims to generalize the findings to other entities. Moreover RE is more attractive in the panel analysis of a sample with large number of entities but short time periods.

## 4. Analysis

Lest a unit root problem, we shall check for stationarity of our input variables. For this study we have chosen the Im, Pesaran and Shin (IPS), which is based on the well-known Dickey-Fuller procedure.

Im, Pesaran and Shin denoted IPS proposed a test for the presence of unit roots in panels that combines information from the time series dimension with that from the cross section dimension, such that fewer time observations are required for the test to have power. Since the IPS test has been found to have superior test power by researchers in economics to analyze long-run relationships in panel data, we will also employ this procedure in this study. IPS begins by specifying a separate ADF regression for each cross-section with individual effects and no time trend:

$$\Delta Z_{it} = \alpha_i + \rho_i Z_{i,t-1} + \sum_{i=1}^{k} \beta_i \Delta Z_{t-i} + e_t \qquad (Eq.7)$$

where $\Delta Z_t$ is the first difference of variable; $\alpha$ is intercept; $\Delta Z_{t-i}$ is lag differences up to lag length *"k"* where k is determined with Schwarz Infomation Criterion (SIC); $e_t$ is White Noise residual term. The IPS hypothesizes the $\rho$ whether it is zero or smaller than zero using separate unit root tests for the *N* cross-section units. Their test is based on the Augmented Dickey-fuller (ADF) statistics averaged across groups. After estimating the separate ADF regressions, the average of the *t*-statistics for $\rho_i$ from the individual ADF regressions, $t_{iTi}(\rho_i)$:

$$\bar{t}_{NT} = \frac{1}{N} \sum_{i=1}^{N} t_{iT}(\rho_i \beta_i) \qquad (Eq.8)$$

The *t*-bar is then standardized and it is shown that the standardized *t-bar* statistic converges to the standard normal distribution as *N* and $T \rightarrow \infty$. Im, Pesaran, and Shin (2003) showed that *t-bar* test has better performance when N and T are small. They proposed a cross-sectionally demeaned version of both test to be used in the case where the errors in different regressions contain a common time-specific component.

The table 4 below presents output of IPS analysis, where all variables appear stationary at level. Equally, the Durbin-Watson values imply that there is no any autocorrelation problem as they are close to 2.

**Table 4.** Output of ADF Analysis

| Variables | ADF | p-value |
|---|---|---|
| ***FDI*** | -16.5742*** | 0.0000 |
| ***Growth*** | -23.8584*** | 0.0000 |
| ***Imports*** | -3.9601*** | 0.0000 |
| ***Exports*** | -3.5712*** | 0.0002 |
| ***Inflation Rate*** | -24.7036*** | 0.0000 |
| ***Interest Rate*** | -3.6988*** | 0.0001 |
| ***EFI*** | -18.0212*** | 0.0000 |

**Notes:** The lag in the table is obtained with Schwarz Information Criterion (SIC) without restricting maximum lag length. The p-values are computed assuming asymptotic normality, and *, **, and *** indicates significance at levels 10%, 5%, and 1% respectively.

### 4.1. Selection of an Appropriate Model

The table 5 presents the results of RFE, BP LM, and Hausman tests which are applied to specify an appropriate model for our panel data. In EFI model, where dependent variable FDI is regressed with independent control variables and EFI, RFE test rejects the null hypothesis which makes the pooled model inappropriate, but the FE model. On the other hand, the BP LM test indicates that the RE model is also appropriate. Indeed, this is a special case where either FE and RE models can be used, or the Hausman test can be utilized to choice one of these two models.

**Table 5.** Model Selection

| ***Redundant Fixed Effect Test*** | ***Breusch-Pagan LM Test*** | ***Hausman Test*** | ***Decision*** |
|---|---|---|---|
| 15.8400<br>(0.0000) | 2090.984<br>(0.0000) | 22.4197<br>(0.0010) | FE model is appropriate. |

**Notes:** In EFI Model the dependent FDI is regressed with control variables (growth, import, export, trade, inflation, and interest rate) and Economic Freedom Index. The null hypothesis of Redundant Fixed Effect Test is no unobserved heterogeneity (no fixed effect) in the model, so pooled model should be used. The null hypothesis for Breusch-Pagan LM Test is no random effect in the model. The null hypothesis for Hausman Test is that there is no correlation between unique errors and the regressors. It implies that both FE and RE estimates are unbiased, but RE is more efficient than FE. So, if null fails to be rejected then RE model would be appropriate. The Redundant Fixed Effect and Breusch-Pagan LM tests show T statistics, whereas the Hausman tests shows Chi-square statistics alongside with probability values in parentheses.

In the decision-making stage, the Hausman test reveals that estimators of RE model are biased. The analysis suggests the FE as an appropriate model for our data. Nonetheless we decide to utilize both models as our sample characteristics make the RE more attractive -countries (entities) in the sample are not functionally identical; purpose of the study is generalizing the findings to other entities too; and large number of entities (156) alongside with short time period (19)- than the FE.

### 4.2. Main Results

The table 6 presents the main results of this study, where findings reveal that EFI has positive and significant impact on FDI under both FE and RE models. The magnitude of this impact considerably changes between two FE models which are estimated with GLS and LSDV methods.

**Table 6.** Results of Panel OLS Models

| | FE (GLS) | FE (LSDV) | RE | Pooled |
|---|---|---|---|---|
| ***C*** | -8.2133*** (0.5070) | -9.6664*** (0.6941) | -8.884*** (0.5960) | -8.0044*** (0.4452) |
| ***Growth*** | 0.0574*** (0.0122) | 0.0786*** (0.0177) | 0.0907*** (0.0175) | 0.1404*** (0.0191) |
| ***Import*** | -1.2719*** (0.0829) | -1.3351*** (0.1105) | -1.1306*** (0.0920) | -0.8171*** (0.0632) |
| ***Export*** | 0.3857*** (0.1207) | 0.2231** (0.0928) | 0.2393*** (0.0787) | 0.2807*** (0.0556) |
| ***Inflation Rate*** | -0.0323** (0.0115) | -0.0316** (0.0165) | -0.0330** (0.0162) | -0.0421*** (0.0171) |
| ***Interest Rate*** | 0.0470** (0.0184) | 0.0395 (0.0298) | 0.0229 (0.0163) | 0.0244 (0.0171) |
| ***EFI*** | 1.1535*** (0.1096) | 1.2653*** (0.1436) | 1.2227*** (0.1291) | 1.1323*** (0.0997) |
| ***Weighted $R^2$*** | 0.6192 | - | 0.1350 | - |
| ***Unweighted $R^2$*** | 0.4934 | 0.4955 | 0.2299 | 0.2413 |
| ***Total Obs.*** | 3432 | 3432 | 3432 | 3432 |
| $\sigma_u$ | - | - | 0.2929 | - |
| $\sigma_e$ | - | - | 0.7071 | - |

**Notes:** The Panel OLS estimation methodology is used to determine FDI with 156 cross-sections and 22 periods (1995-2016). The FE (GLS - Generalized Least Squares) model has fixed effect of cross-section and GLS cross-section weights; FE (LSDV - Least Squares with Dummy Variables) model has fixed effect of cross-section without any GLS weights; RE model has random effect of cross-section; and Pooled model is free of any effect specifications and GLS weights. The White standard errors are presented in parentheses, and *, **, and *** denote significance at 10%, 5%, and 1% levels respectively. The $\sigma_u$ and $\sigma_e$ represent Swamy and Arora estimator of variance components of random effect (cross-section and idiosyncratic respectively) with rho numbers.

More specifically, the $R^2$ of FE (GLS) model implies that 62% of variation in FDI is accounted by EFI and control variables. Meantime all variables appear statistically significant between at 1%-5% levels. The coefficients of GDP growth, Export, Interest, and EFI respectively as 0.0574, 0.3857, 0.0470, and 1.1535 reveal that these variables have positive impact on FDI. Particularly, the role of control variables is blatant. The greatest impact on FDI appears as Imports, Exports, and EFI. A unit increase in imports shrinks FDI by 1.27 units, whereas a unit increase in exports and EFI augments FDI by 0.38 units and 1.15 units respectively. The roles of inflation and interest rates also make sense as a unit increase of inflation diminishes FDI by 0.03 and a unit increase in interest enhances FDI by 0.05 units respectively.

In the FE (LSDV) model $R^2$ drops to 50 % and the coefficient of EFI increases to 1.26 preserving its significance at 1% level. Equally, all variables except interest preserve their significances while coefficients of export and interest decrease, and coefficients of growth and imports increase. The coefficient of inflations appears approximately same as in FE (GLS) case. Similar scenario repeats in RE and Pooled models as well. However, the $R^2$ values in the RE and pooled models have got even worse as they drop to 23% and 24 % respectively. The estimates of EFI in RE and Pooled models indicate that a unit increase in EFI augments FDI by 1.22 units and 1.13 units respectively. Similar to FE (LSDV) case, both in RE and Pooled models interest rates appear to be statistically insignificant.

To conclude, we rely on estimates of Fixed models as model selection tests in table 5 points out that RE model is biased, so FE is appropriate. Therefore, results of both FE models GLS and LSDV are valid, however the GLS performs better as it accounts much greater variations of FDI.

**4.3. Regional Results**

The table 7 presents regional results of our panel study analysis where pooled, FE, and RE models are used on the basis of aftermath of model selection test such as RFE, BP LM, and Hausman. In case of European (EU) sample, FE model appears more appropriate where 54.17% of variation in FDI is accounted by control variables and EFI. According to estimated results, EFI gets coefficient of 2.40 which is significant at 1% level. It indicates that a unit increase in EFI augments FDI by almost two and half folds. Indeed this is the highest magnitude of EFI among all studied regions. Meantime, GDP growth, export, and interest rate variables also get positive estimates which are significant at 1% level. A unit increase in these three variables augments FDI by 0.06 units, 1.06 units, and 0.27 units respectively. The positive relationships between growth-FDI and interest rates-FDI make sense from foreign investors' perspective as they rely on growing economy and love high interest returns.

**Table 7.** Results of Regional Panel OLS Analysis

| | EU | AS | AF | NA | LA | OC | FC | SS | PS |
|---|---|---|---|---|---|---|---|---|---|
| **C** | -9.9702*** | -3.5245*** | -8.7198*** | -10.0040*** | -3.4354*** | -4.6717** | -7.2338*** | -8.2462*** | -4.5898*** |
| | (1.2478) | (0.9695) | (1.2369) | (2.8418) | (1.0384) | (2.2399) | (1.6575) | (1.2205) | (1.4618) |
| **Growth** | 0.0612*** | 0.0341*** | 0.1035*** | 0.0445*** | 0.0474*** | 0.2057*** | 0.2200*** | 0.1432*** | 0.0800*** |
| | (0.0216) | (0.0029) | (0.0303) | (0.0074) | (0.0056) | (0.0545) | (0.0480) | (0.0331) | (0.0310) |
| **Import** | -1.3954*** | -0.8771*** | -1.5795*** | -1.4940*** | 0.4727 | -0.6969*** | -0.6319*** | -1.2972*** | -0.8827*** |
| | (0.2539) | (0.1424) | (0.1466) | (0.2981) | (0.3899) | (0.2971) | (0.2281) | (0.1399) | (0.2052) |
| **Export** | 1.0613*** | 0.5581*** | 0.2779** | 0.1311 | 0.3262*** | 0.0961 | 0.5733*** | 0.2792** | 0.2473 |
| | (0.2061) | (0.1624) | (0.1309) | (0.2342) | (0.0819) | (0.2378) | (0.1378) | (0.1338) | (0.1674) |
| **Inflation Rate** | -0.0782*** | -0.0881*** | 0.0235 | -0.0531** | -0.2219** | -0.2894*** | 0.0145 | -0.2321*** | -0.0203 |
| | (0.0256) | (0.0337) | (0.0222) | (0.0264) | (0.1027) | (0.0299) | (0.0442) | (0.0334) | (0.0416) |
| **Interest Rate** | 0.2752*** | 0.2220*** | 0.1654*** | 0.2855** | 0.1889 | -0.0186 | 0.0649 | -0.0232 | 0.0614** |
| | (0.0378) | (0.0266) | (0.0417) | (0.1389) | (0.1604) | (0.0811) | (0.0620) | (0.0729) | (0.0201) |
| **EFI** | 2.4039*** | 1.3934*** | 1.2105*** | 1.9090*** | 1.3933*** | 0.7878** | 0.7029** | 0.7776*** | 0.9873*** |
| | (0.2429) | (0.2275) | (0.2976) | (0.6369) | (0.3322) | (0.3510) | (0.3470) | (0.2914) | (0.2940) |
| Weighted $R^2$ | 0.5417 | 0.7905 | 0.5298 | 0.2868 | 0.6677 | 0.2228 | 0.1508 | 0.5712 | 0.1508 |
| Unweighted $R^2$ | 0.5129 | 0.5969 | 0.4474 | 0.3580 | 0.5327 | 0.3213 | 0.2114 | 0.4899 | 0.0642 |
| Cross-section | 42 | 26 | 45 | 12 | 15 | 15 | 23 | 40 | 15 |
| Total Obs | 924 | 572 | 990 | 264 | 330 | 330 | 506 | 880 | 330 |
| $\sigma_u$ | - | - | - | 0.2881 | - | 0.1414 | 0.1318 | - | 0.3984 |
| $\sigma_e$ | - | - | - | 0.7119 | - | 0.8586 | 0.8682 | - | 0.6016 |
| RFE | 5.8565 | 17.9109 | 12.9808 | 13.7719 | 25.1884 | 12.5855 | 10.6883 | 12.4504 | 17.8125 |
| | (0.0000) | (0.0000) | (0.0000) | (0.0000) | (0.0000) | (0.0000) | (0.0000) | (0.0000) | (0.0000) |
| BP LM | 172.7284 | 174.9032 | 308.2883 | 63.7747 | 126.9496 | 35.0050 | 51.2183 | 268.7579 | 319.2704 |
| | (0.0000) | (0.0000) | (0.0000) | (0.0000) | (0.0000) | (0.0000) | (0.0000) | (0.0000) | (0.0000) |
| Hausman | 16.3472 | 19.7777 | 22.8379 | 8.4933 | 59.8990 | 12.1062 | 7.2847 | 20.8435 | 10.9184 |
| | (0.0120) | (0.0030) | (0.0009) | (0.1076) | (0.0000) | (0.0596) | (0.2953) | (0.0020) | (0.0909) |
| Decision | FE | FE | FE | RE | FE | RE | RE | FE | RE |

**Notes:** The Panel OLS estimation methodology is used to determine FDI in Europe (EU), Asia (AS), North Africa (AF), North America (NA), Latin America (LA), Oceania (OC), Fragile-Conflict States (FC), Sub-Saharan (SS), and Post-Soviet (PS) countries with EFI and control variables over 22 periods (1995-2016). Each sample is estimated on its appropriate decided model where FE and RE has cross-section fixed and random effects respectively, whereas pooled model is free of any effect specifications. The null hypotheses of Redundant Fixed Effect (RFE), Breusch-Pagan (BP) LM , and Hausman tests are presented in table 5. The Redundant Fixed Effect and Breusch-Pagan LM tests show T statistics, whereas the Hausman tests shows Chi-square statistics alongside with probability values in parentheses. The White standard errors are presented in parentheses, and *, **, and *** denote significance at 10%, 5%, and 1% levels respectively. The $\sigma_u$ and $\sigma_e$ represent Swamy and Arora estimator of variance components of random effect (cross-section and idiosyncratic respectively) with rho numbers.

On the other hand, import and inflation rate variables get negative coefficients which imply inverse relationship to FDI. A unit increase in these variables shrinks FDI by 1.39 units and 0.08 units respectively. The negative impact of inflation rates on FDI makes sense as foreign investors avoid inflationary markets rather prefer price level to be stable in long-term in order to initiating or continuing their investment activities in that country.

For Asian (AS) sample, the model selection tests again suggest FE model as the most appropriate one where 79% of FDI is explained by input variables. The results of this model appear similar to European case. Here, EFI generates a coefficient of 1.39 at 1% significance level indicating that a unit increase in EFI enhances FDI by 1.39 units. This impact was twofold in European case. This might be due to high levels of democracy (in terms of protection of investors' rights) and well-established financial markets in European markets comparing to Asian markets. Besides, the positive coefficients of Growth, Export, and Interest rate that a unit increase in these variables scales FDI by 0.03 units, 0.56 units, and 0.22 units respectively.

We consider two different samples for African countries: North African (AF) and Sub-Saharan (SS). Because these two regions have significant differences in market characteristics. For AF sample, the model selection test points FE as appropriate model which account nearly 53% of variations in FDI. The results estimate positive impacts running from GDP growth, Exports, Interest rates, and EFI to FDI, whereas imports generate a negative impact. Here, a unit increase in GDP growth, Exports, Interest Rates augments FDI by 0.10 units, 0.28 units, and 0.16 units respectively. FDI is highly sensitive to the trade activities in African regions; therefore the model derives a significant coefficient for exports (0.28) and imports (-1.58). The inflation rates get a positive estimate, however, it fails to be statistically significant. Additionally, notice that interest rates derive positive and statistically significant, however, its magnitude is smaller than EU and AS ones. This indicates that interest rates have limited impact on attracting foreign investors. The same scenario presents in EFI factor. It takes a coefficient of 1.21 which is significant at 1% level, but its magnitude is smaller than EU and AS. One can argue that international investors are less sensitive to activates of weakly-established financial markets, thus, this might be a key reason behind these results.

Interestingly, the model selection tests suggest RE as the most appropriate model for North American (NA) sample and FE model for Latin American (LA) sample. Although all series except exports appear to be statistically significant in NA, RE model accounts only 36% variation in FDI. Similar to EU and AS cases, RE model estimates positive coefficient for GDP growth, exports, interest rates, and EFI, whereas the model estimates negative coefficient for imports and inflation rates. On the other hand, FE model accounts 66% of variation in FDI by generating statistically significant coefficient for GDP growth, Exports, Inflation rates, and EFI.

Notice that magnitude of GDP growth variable in both NA and LA is almost equal, around 0.04. However, there are remarkable differences in magnitudes and in significance levels of other explanatory variables of NA and LA. For instance, reaction of FDI to inflation rates is more aggressive in case of LA where a unit increase in inflation decreases FDI by 0.22 units. It might be due proneness of inflation of LA region which likely to deter foreign investments. More importantly, the EFI coefficient implies that economic freedom has a greater value in NA region in terms of attracting FDI comparing to LA one. Once again it shed light on structural differences between financial markets of these two regions.

Unquestionably the most interesting parts of this study are the cases of Oceanian (OC), Fragile-Conflict (FC), Sub-Saharan (SS), and Post-Soviet (PS) states. To our knowledge these regions are either never or rarely analyzed. Therefore with this study we fill this room. The aftermath of analysis shows that in cases of OC, FC, and PS, the RE model appears as the most appropriate and it accounts around 32%, 21%, and 6% of variation in FDI respectively. In case of SS, model selection test suggest FE model which accounts 57% of variation in FDI. Moreover, in all these cases the EFI generates positive and significant coefficients. This significance is relatively lesser in cases of OC and FC than others. More interestingly, coefficients of EFI get value of 0.70, 0.77, 0.79, and 0.99 in FC, SS, OC, and PS respectively. Notice that these values are the weakest (the lowest) ones among other samples' EFI coefficients. The weakest EFI appears in FC region which is followed by SS. This implies that the impact of EFI on FDI is lesser comparing to others regions, but it is still significant factor in determination of FDI.

The control variables of GDP growth, Imports, and exports also contribute significance to these models. Especially, imports and GDP growth appear as locomotives of FDI as they generates significant coefficients of -0.70 and 0.20 in OC, -0.63 and 0.22 in FC, -1.29 and 0.14 in SS, and -0.88 and 0.08 in PS. As well as, the role of inflation rates in OC and SS is also remarkable. It implies that a unit increase in inflation rates shrink in FDI by 0.29 units in OC, and 0.23 units in SS zone.

## 5. Conclusion

The study investigates the impact of economic freedom on FDI inflows globally taking into account often neglected regions such as Sub-Saharan, Post-Soviet, and Conflict-Affected countries in a panel data. The global analysis shows that FDI is largely affected by internal-external trades and economic freedom of the countries in FE model. Although RE model generates quite similar results, Hausman test implies that they are biased.

On the regional basis, the study perfectly demonstrates remarkable impact of EFI on FDI as in all cases it generates positive estimates at 1% significance level, except OC and FC samples where the significance is limited with 5% level. More interestingly the sensitivity of FDI on EFI is relatively less in OC, FC, and SS states, and high in EU, NA, and AS regions. The analysis derives significant coefficients for economic freedom variable, but indifferent in magnitudes. In European sample EFI obtains the largest magnitude with FE model where a unit increase in EFI augments FDI by 2.40 units. European sample is followed by North American sample with EFI value of 1.90; Asian and Latin American sample with EFI values of 1.39; North African sample with EFI value of 1.21; Post-Soviet sample with 0.99; Ocenian and Sub-Saharan samples with EFI value of 0.78; and Fragile-Conflict affected states sample with lowest EFI value of 0.70.

As a result, it suggest to pursue EFI-oriented policy and to implement new plans to attract more and more FDI which will bring in new innovative and automation based technologies that can rejuvenate the host country's existing manufacturing base. Furthermore, human labour transfer in the form of highly skilled, experienced and knowledge-versed is a remarkable move to boost the country's economic growth. As a twin opportunity, the results also suggest to implement a trade regime that encourage domestic manufacturer by various of policies including weakening in value of national monetary unit (exchange rates) in order to create an export-oriented industries in the economy. But while implementing all these policies, the economy should first of all maintain secure and stable financial grounds with high standards of liberal economic regulations.

The control variables of GDP growth, Imports, and exports also plays significant role in determination of FDI. This is more blatant in well-established (developed) financial markets such as EU, AS, and NA countries; whereas significance of control variables remain very limited in weakly-established or restricted financial markets such as OC, FC, SS, and PS countries. It might be motivated by high levels of political and financial uncertainty in these regions. To make it clear, we suggest further researchers to consider impacts of other factors such as political instability, corruption level, institutional rights, financial and labor market regulations, and credit default risks.

## References

Ajide, K.B., and Eregha, P.B. (2014). Economic Freedom and Foreign Direct Investment in Ecowas Countries: A Panel Data Analysis. *Applied Econometrics and International Development Journal,* 14 (2), 163-174.

Asiedu, E. (2006). Foreign Direct Investment in Africa: The Role of Natural Resources, Market Size, Government Policy, Institutions and Political Instability. *The World Economy,* 29 (1), 63-77.

Bell, A.J.D., and Jones, K. (2015). Explaining Fixed Effects: Random Effects modelling of Time-Series Cross-Sectional and Panel Data. *Political Science Research and Methods*, 3(1), 133-153.

Bengoa, M., and Sanchez-Robles, B. (2003). Foreign Direct Investment, Economic Freedom and Growth: New Evidence from Latin America. *European Journal of Political Economy,* 19 (3), 529-545.

Bondell, H.D., Krishna, A., and Ghosh, S.K. (2011). Joint Variable Selection for Fixed and Random Effects in Linear Mixed-Effects Models. *Biometrics*, 66 (4), 1069-1077.

Borensztein, E., De Gregorio, J. & Lee, J.W. (1998). How Does Foreign Direct Investment Affect Economic Growth? *Journal of International Economics,* 45 (1), 115– 135.

Carkovic, M., and Levine, R. (2005). Does Foreign Direct Investment Accelerate Economic Growth?. In T. Moran, E.Graham, & M. Blomstrom (eds) *"Does Foreign Direct Investment Promote Development?"* (pp. 195-220). Center for Global Development and Institute for International Economics, Washington DC.

Chaib, B., and Siham, M. (2014). The Impact of Institutional Quality in Attracting Foreign Direct Investment in Algeria.*Topics in Middle Eastern and African Economies,* 16 (2), 140-163.

Clark, T.S., and Linzer, D.A. (2014). Should I Use Fixed or Random Effects? *Political Science Research and Methods*. March, 1-10. doi:10.1017/psrm.2014.32

Fofana, M.F. (2014).The Influence of Measures of Economic Freedom on FDI: A Comparison of Western Europe and Sub-Saharan Africa. *Global Economy Journal,* 14 (3-4), 399–424.

Hornberger, K., Battat, J., and Kusek, P. (2011). *Attracting FDI: How Much Does Investment Climate Matter?* New York: The World Bank Group.

Im, K.S., Pesaran, M.H., and Shin, Y. (2003). Testing for Unit Roots in heterogeneous Panels. *Journal of Economics*, 115, 53-74.

Kinney S.K., and Dunson, D.B. (2007). Fixed and Random Effects Selections in Linear and Logistic Models. *Biometrics*, 63 (3), 690-698.

Kudaisi, B.V. (2014). An Empirical Determination of Foreign Direct Investment in West Africa Countries: A Panel Data Analysis. *International Journal of Development and Economic Sustainability,* 2(2), 19-36.

Lund, M.T. (2010). *Foreign Direct Investment: Catalyst of Economic Growth?* Ph.D. Dissertation, University of Utah Graduate School.

Mohamed, S.E., and Sidiropoulos, M.G. (2010). Another Look At The Determinants of Foreign Direct Investment in Mena Countries: An Empirical Investigation. *Journal of Economic Development,* 35 (2), 75-95.

Park, H.M. (2009). Linear Regression Models for Panel Data Using SAS, Stata,LIMDEP, and SPSS. Linear Regression Models for Panel Data. The Trustees of Indiana University.

Pearson, D., Nyonna, D., and Kim, K.J. (2012). The Relationship between Economic Freedom, State Growth and Foreign Direct Investment in US States. *International Journal of Economics and Finance,* 4 (10), 140-146.

Quazi, R. (2007). Economic Freedom and Foreign Direct Investment in East Asia. *Journal of the Asia Pacific Economy,* 12 (3), 329–344.

Sambharya, R., and Rasheed, A. (2015). Does Economic Freedom in Host Countries Lead to Increase Foreign Direct Investment? *Competitiveness Review,* 25 (1), 2-24.

## Appendix

**Table 8.** Sampled Countries

| SAMPLED COUNTRIES | | SAMPLED COUNTRIES | |
|---|---|---|---|
| 1 | Albania | 79 | Latvia |
| 2 | Algeria | 80 | Lebanon |
| 3 | Angola | 81 | Lesotho |
| 4 | Argentina | 82 | Libya |
| 5 | Armenia | 83 | Lithuania |
| 6 | Australia | 84 | Luxembourg |
| 7 | Austria | 85 | Macedonia |
| 8 | Azerbaijan | 86 | Madagascar |
| 9 | Bahrain | 87 | Malawi |
| 10 | Bangladesh | 88 | Malaysia |
| 11 | Barbados | 89 | Mali |
| 12 | Belarus | 90 | Malta |
| 13 | Belgium | 91 | Mauritania |
| 14 | Belize | 92 | Mauritius |
| 15 | Benin | 93 | Mexico |
| 16 | Bolivia | 94 | Moldova |
| 17 | Bosnia and Herzegovina | 95 | Mongolia |
| 18 | Botswana | 96 | Morocco |
| 19 | Brazil | 97 | Mozambique |
| 20 | Bulgaria | 98 | Namibia |
| 21 | Burkina Faso | 99 | Nepal |
| 22 | Burundi | 100 | New Zealand |
| 23 | Cabo Verde | 101 | Nicaragua |
| 24 | Cambodia | 102 | Niger |
| 25 | Cameroon | 103 | Nigeria |
| 26 | Canada | 104 | Norway |
| 27 | Central African Republic | 105 | Oman |
| 28 | Chad | 106 | Pakistan |
| 29 | Chile | 107 | Panama |
| 30 | China | 108 | Papua New Guinea |
| 31 | Colombia | 109 | Paraguay |
| 32 | Costa Rica | 110 | Peru |
| 33 | Cote d'Ivoire | 111 | Philippines |
| 34 | Croatia | 112 | Poland |
| 35 | Cuba | 113 | Portugal |
| 36 | Cyprus | 114 | Qatar |
| 37 | Czech Republic | 115 | Republic of Congo |

| | |
|---|---|
| 38 | Democratic Republic of Congo |
| 39 | Denmark |
| 40 | Djibouti |
| 41 | Dominican Republic |
| 42 | Ecuador |
| 43 | Egypt |
| 44 | El Salvador |
| 45 | Equatorial Guinea |
| 46 | Estonia |
| 47 | Ethiopia |
| 48 | Fiji |
| 49 | Finland |
| 50 | France |
| 51 | Gabon |
| 52 | Georgia |
| 53 | Germany |
| 54 | Ghana |
| 55 | Greece |
| 56 | Guatemala |
| 57 | Guinea |
| 58 | Guinea-Bissau |
| 59 | Guyana |
| 60 | Haiti |
| 61 | Honduras |
| 62 | Hong Kong |
| 63 | Hungary |
| 64 | Iceland |
| 65 | India |
| 66 | Indonesia |
| 67 | Iran |
| 68 | Ireland |
| 69 | Israel |
| 70 | Italy |
| 71 | Jamaica |
| 72 | Japan |
| 73 | Jordan |
| 74 | Kazakhstan |
| 75 | Kenya |
| 76 | Kuwait |
| 77 | Kyrgyz Republic |
| 78 | LAOS |
| 116 | Romania |
| 117 | Russian |
| 118 | Rwanda |
| 119 | Saudi Arabia |
| 120 | Senegal |
| 121 | Sierra Leone |
| 122 | Singapore |
| 123 | Slovakia |
| 124 | Slovenia |
| 125 | South Africa |
| 126 | South Korea |
| 127 | Spain |
| 128 | Sri Lanka |
| 129 | Suriname |
| 130 | Swaziland |
| 131 | Sweden |
| 132 | Switzerland |
| 133 | Syria |
| 134 | Tajikistan |
| 135 | Tanzania |
| 136 | Thailand |
| 137 | The Bahamas |
| 138 | The Gambia |
| 139 | The Netherlands |
| 140 | Togo |
| 141 | Trinidad and Tobago |
| 142 | Tunisia |
| 143 | Turkey |
| 144 | Turkmenistan |
| 145 | Uganda |
| 146 | Ukraine |
| 147 | United Arab Emirates |
| 148 | United Kingdom |
| 149 | United States |
| 150 | Uruguay |
| 151 | Uzbekistan |
| 152 | Venezuela |
| 153 | Vietnam |
| 154 | Yemen |
| 155 | Zambia |
| 156 | Zimbabwe |